# The Read-Out Shutter Unit of the Euclid VIS Instrument

L. Genolet[*,a], E. Bozzo[a], S. Paltani[a], N. Autissier[b], C. Larcheveque[b], C. Thomas[b]

[a]Department of Astronomy, University of Geneva, Chemin d'Ecogia 16, 1290, Versoix, Switzerland;
[b]APCO Technologies, Chemin de Champex 10, CH-1860 Aigle, Switzerland.

## ABSTRACT

Euclid is the second medium-size mission (M2) of the ESA Cosmic Vision Program, currently scheduled for a launch in 2020. The two instruments on-board Euclid, VIS and NISP, will provide key measurements to investigate the nature of dark energy, advancing our knowledge on cosmology. We present in this contribution the development and manufacturing status of the VIS Read-out Shutter Unit, whose main function is to prevent direct light from falling onto the VIS CCDs during the read-out of the scientific exposures and to allow the dark-current/bias calibrations of the instrument.

## 1. INTRODUCTION: THE EUCLID-VIS INSTRUMENT

Euclid-VIS is the visible imager on-board the Euclid space mission, which was selected by ESA on October 2011 as the M2 mission of the Cosmic Vision program to be launched in 2020. Together with NISP, that complements the mission payload and performs observations in the near-infrared domain, VIS will provide key measurements through which Euclid will be able to infer the nature of dark energy and quantify its role in the evolution of the Universe. The VIS instrument consists of: (i) the Focal Plane Assembly (FPA), which is responsible for the detection of visible light and its conversion to an electrical signal; (ii) the Calibration Unit (CU), which illuminates the FPA with the use of LEDs to perform flat-field calibrations; (iii) the Readout Shutter Unit (RSU), blocking the VIS optical path during the read-out of the science exposures and dark-current/bias calibrations; (iv) the Control and Data Processing Unit (CDPU), which controls the instrument, performs the processing of the VIS images and transfers the data packets to the spacecraft; (v) the Power and Mechanism Control Unit (PMCU), which controls the RSU, the CU and monitors the FPA temperature sensors.

Since the initial stages of the Euclid mission phases, the University of Geneva (Switzerland) has been in charge of the design and development of the RSU. The design of the unit has been consolidated during phases A and B with the support of industry (Ruag Space, Switzerland). Following the successful completion of the instrument preliminary design review at the end of phase B, the further development, detailed design, and manufacturing of the RSU up to the delivery of the flight model is being carried out by the Swiss company APCO Technologies. We present in Sect. 2 the overall and detailed design of the unit that has been demonstrated to successfully fulfill the imposed requirements (Sect. 2.1). We highlight some particularly challenging aspects of the RSU design in Sect. 3, and provide a summary of the most recently encountered issues with the RSU manufacturing and test in Sect. 4. The RSU model philosophy, as well as the updated status of the unit development, is provided in Sect. 5. Our conclusions are reported in Sect. 6.

---

[*]ludovic.genolet@unige.ch

## 2. RSU DESIGN DESCRIPTION

The main function of the RSU is to prevent the science beam from the telescope to arrive onto the focal plane of the VIS instrument, while the latter is reading out the CCDs at the end of each scientific exposure. To fulfill its function, the RSU is equipped with a hatch, which occults the surface area of the focal plane, and a stepper motor, which is used to drive the rotation of the hatch between the open and closed position.

### 2.1 Key requirements

There are a number of requirements that make the development of the RSU particularly challenging. The first one is related to the high spacecraft stability that is needed to achieve the VIS instrument pointing accuracy. As the RSU is operated during the scientific observations and its motion (~10 s) is too short to be actively compensated through the usage of the spacecraft altitude and orbit control system (reaction time ~60s), it is mandatory that the design of the unit minimizes the de-pointing of the satellite and the excitation of its critical vibration modes. This is achieved mainly through the use of a flywheel and the accurate balancing of the different moving RSU sub-assemblies, adjusting their center of gravity and moment of inertia with an exquisite precision during the manufacturing process and the assembly. It has been shown by APCO that the improved RSU design in phase C largely damps the exported torque of the unit and decreases the resulting micro-vibrations to an acceptable level (see Sect. 3.2).

The other requirements concern the life-time of the unit and its operations in a strictly controlled environment. As the Euclid mission has to be operated for more than 6 years in order to achieve its science goals ([1]), it is expected that the RSU will perform during this period more than 350'000 opening and closing cycles at 110-160 K. The RSU operations at these cold temperatures are required due to the proximity of the shutter to the FPA that is equipped with CCDs reaching the highest performances only at cold temperatures ([2]). The presence of this critical optical surface poses also strict requirements to the RSU on the cleanliness and stray-light level, which should be kept low during the entire motion of the relatively large RSU leaf. The size of this part, as well as the overall RSU mass and volume, is mainly driven by the need to occult the light from the telescope, which corresponds to a rectangular beam of 250x275 mm. Finally, it is worth highlighting that for tests and integration activities, the RSU is also requested to be fully functional at room temperature.

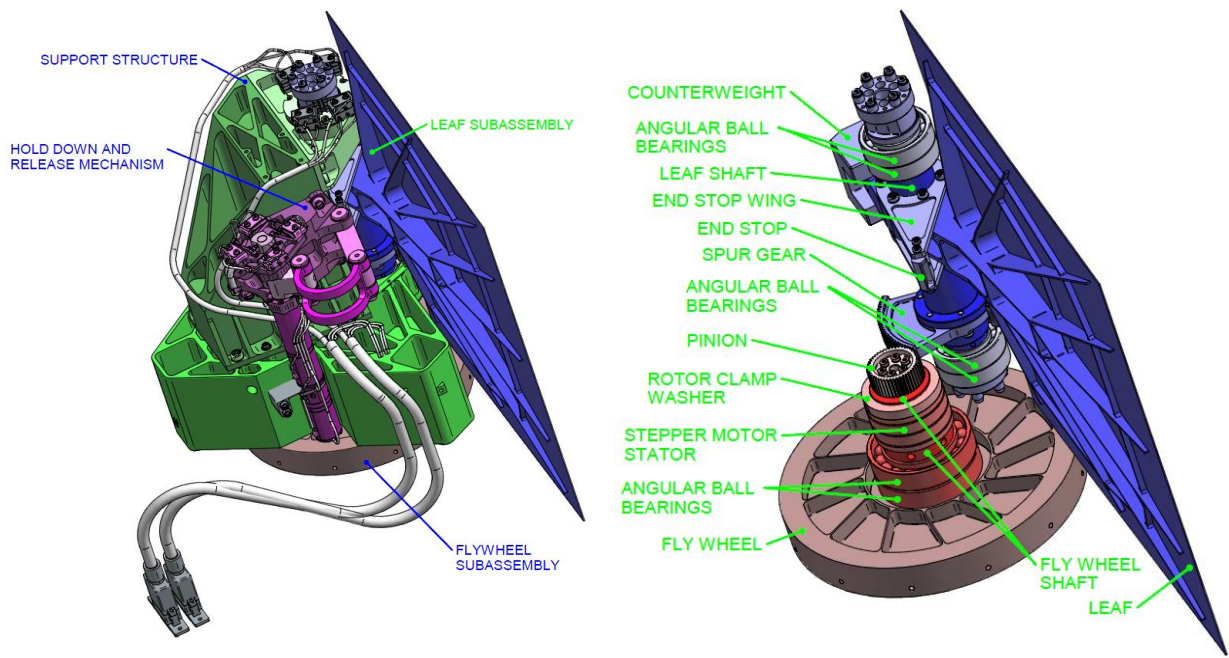

**Figure 2-1 Left: general view of the RSU showing the location of the main unit sub-assemblies. Right: Details of the leaf and flywheel subassembly.**

## 2.2 Detailed RSU design

The RSU consists of 4 sub-systems: the leaf subassembly (LS), the flywheel subassembly (FwS), the hold down and release mechanism (HDRM) and the support structure. The main function of the LS is to intercept the optical science beam when the RSU is in closed position. The LS consists of an aluminum leaf mounted on a shaft held by two pairs of angular ball bearings in face to face arrangement. A counterweight made of tungsten was introduced to balance the LS around the rotation axis. A spur gear is used to connect the LS and the FwS. A rotation of 70° is necessary to move the LS from its opened position to its closed position.

The HDRM function is to lock the RSU during launch and definitively unlock the LS once in orbit. The subassembly was designed to provide also the possibility to reversibly lock and unlock the LS during on-ground tests. The HDRM consists of a toggle mechanism comprising the motor arm, the compliant arm, the hook, and the base plate. It is driven by a DC motor with built-in redundancy. In locked position, the toggle mechanism applies a preload onto the LS end stop. The compliant arm has a significantly lower stiffness than the 3 other arms and it is specifically designed to deliver the adequate preload thanks to its bending during the locking sequence. By design, the preload in the locked position is maintained by the toggle mechanism and not by the motor. The latter is used only to move the toggle mechanism between the locked and the unlocked positions.

The FwS consists of the flywheel, mounted on the corresponding shaft and held by a pair of angular ball bearing in back-to-back arrangement. A direct-drive stepper motor, comprising a stator installed on the "base plate 1" and a rotor mounted on the flywheel shaft, drives the motion of the FwS and the LS through the LS pinion (which is meshed with the FwS spur gear). The selected stepper motor for the RSU is developed and manufactured by SAFRAN-SAGEM. It is able to perform 360 steps per revolution and is equipped with redundant windings. The main function of the FwS is to balance the angular momentum generated by the LS rotation, decreasing the exported torque and micro-vibrations induced by the RSU on the spacecraft.

Cold and warm redundant end-switches are installed on the tip of the LS shaft. The end switches are actuated by a cam pressing on a roller, which bends a lever. These switches are used to verify that the LS is either in the open or the closed position during normal operations. Similar switches are also mounted close to the HDRM in order to verify the successful release of the RSU leaf once in orbit. As the HDRM has to be operated only once after launch, cold redundant switches are implemented for this subassembly.

A sketch of the RSU in open and closed position is shown in Figure 2-3.

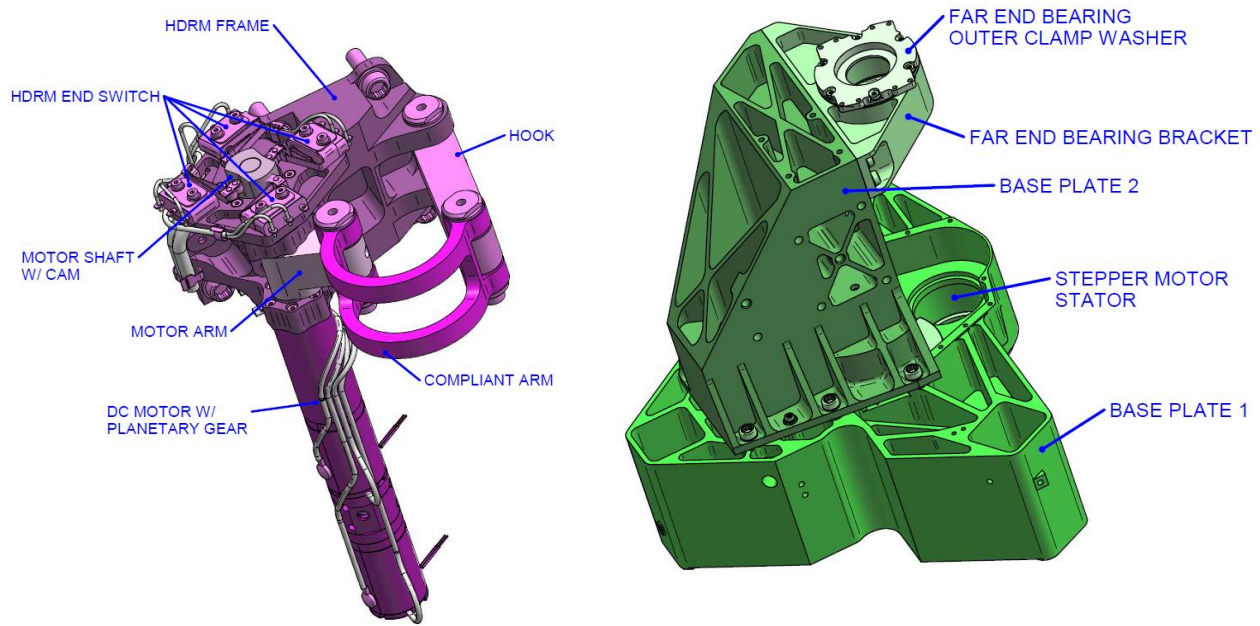

**Figure 2-2 Left: the hold-down and release mechanism. Right: the support structure.**

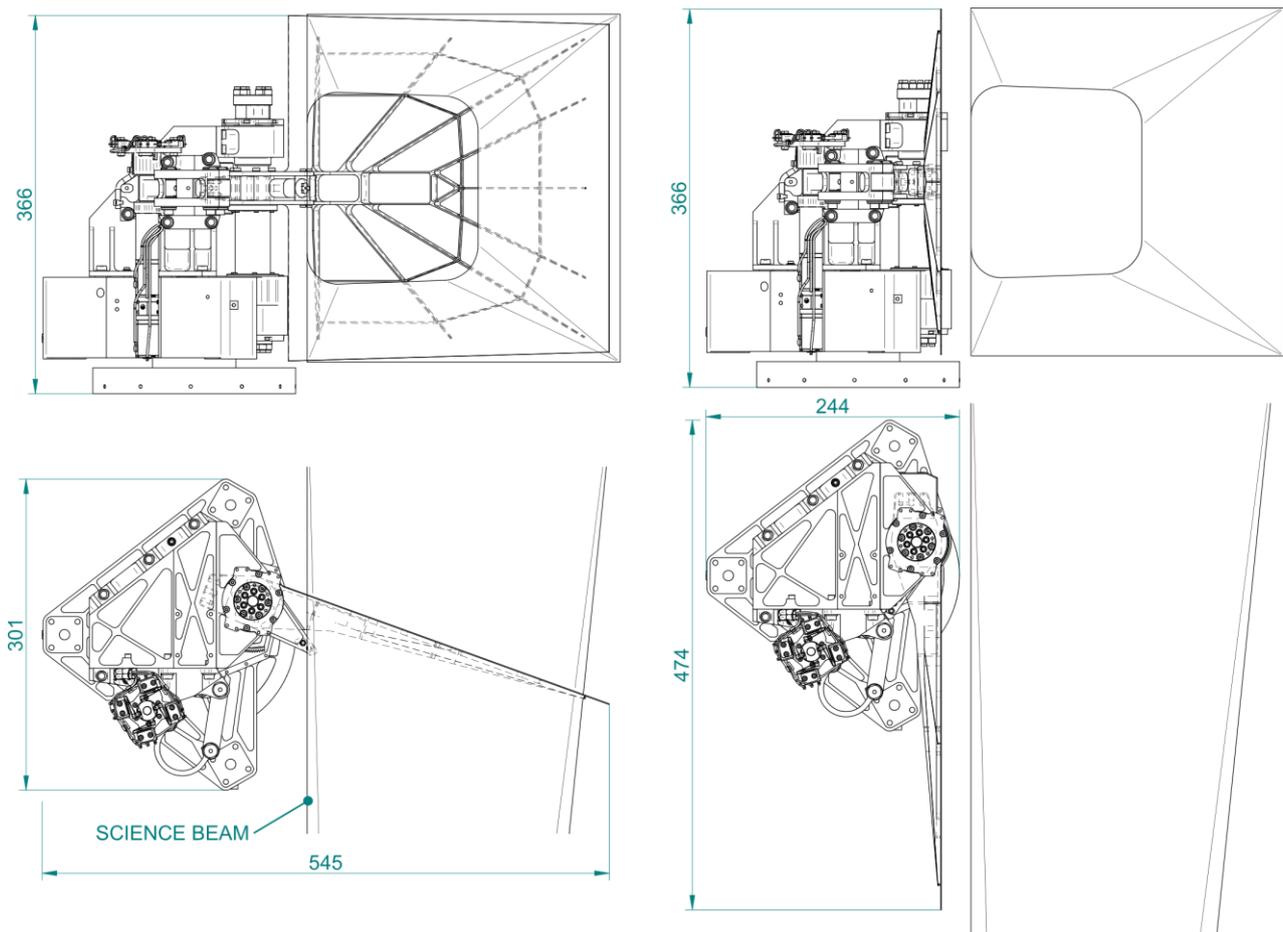

**Figure 2-3 Left sketches: RSU in closed position. Right sketches: RSU in open position. The optical science beam is indicated in all sketches.**

## 3. SPECIFIC RSU DESIGN CHALLENGES

### 3.1 De-pointing

The motion of the LS generates an angular momentum, and thus a de-pointing on the spacecraft, that is compensated by the motion of the FwS in the opposite direction. A gear ratio of 3 between the FwS and the LS is implemented in order to improve the achievable resolution in the position of the LS and decrease by a factor of 3 the inertia of the FwS with respect to that of the LS (in turns optimizing the mass of the RSU). The residual de-pointing left after this corrective measure is mainly due to:

- The inertia mismatch between the LS and the FwS. A custom moment of inertia test stand has been developed by Schenck S.A.S for the Euclid RSU project in order to comply with the challenging low admissible mismatch. This machine can measure the moment of inertia of the LS and FwS down to an accuracy of 0.1%. Adjustable counterweights have been implemented on the FwS in order to optimize the inertia after the measurements, and the materials of the RSU have been selected in order to minimize their relative thermal expansions/contractions in the entire temperature operational range of the unit.
- Static imbalance due to the offset of the center of gravity (CoG) of the different RSU parts with respect to the main rotation axis. This off-set gives rise to a rotating force perpendicular to the rotation axis and thus affects the Euclid de-pointing performance. To minimize this effect, APCO is planning an iterative sequence of measurements and re-machining of the different RSU parts during the final stages of the unit assembly.

- Angular displacement of the moving sub-assemblies of the RSU. The de-pointing is proportional to the angular displacement of the different RSU parts during the motion of the leaf. The design and the angular displacement of the RSU are constrained by the vicinity of optical components, limiting the volume allocated to the RSU.
- The gear backlash. Spur gears, as the one implemented in the RSU, requires a backlash to work properly, but the latter was shown to significantly affect the de-pointing performances. Due to the backlash, the FwS angular momentum generated in the initial stages of the RSU actuation is indeed not counter-balanced by the LS rotation until the pinion of the FwS gets in proper contact with the LS gear. In order to minimize this effect, it is foreseen to limit as much as possible the spur gear backlash and shorten the initial uncompensated motion of the FwS.
- Parallelism between the 2 rotation axes of the sub-assemblies. This is currently being taken care of by using tolerances in the manufacturing procedure of the different RSU parts and will be improved through the machining of parts when assembled.
- Cross-terms of the inertia matrix (products of inertia). It has been shown during phase C that the rotating sub-assemblies of the RSU should be symmetrically designed around the rotation axis (i.e. the z-axis in the equation below) in order to avoid any residual angular momenta around the correspondingly orthogonal directions (i.e. $L_x$, $L_y$ in the equation below):

$$\begin{pmatrix} I_{xx} & I_{xy} & I_{xz} \\ & I_{yy} & I_{yz} \\ & & I_{zz} \end{pmatrix} \begin{pmatrix} 0 \\ 0 \\ \dot{\theta}_z \end{pmatrix} = \begin{pmatrix} I_{xz} \\ I_{yz} \\ I_{zz} \end{pmatrix} \dot{\theta}_z = \begin{pmatrix} L_x \\ L_y \\ L_z \end{pmatrix}$$

As it can be seen from the equation above, a symmetric design of the different parts around the rotation axis naturally minimizes the cross-terms of the inertia matrix: $I_{xz}$, $I_{yz}$, $I_{zx}$, and $I_{zy}$. The remaining parasitic product of inertia will be corrected by dynamic balancing of the LS.

## 3.2 Micro-vibration

As the leaf occulting the science beam is relatively large, a significant number of structural modes could be excited starting from 130 Hz, giving rise to additional angular momenta (note that the 1st RSU mode is at 134 Hz). The RSU stepper motor is the main source of excitation of these modes. Stepper motors have a long heritage in space projects, but one of their disadvantages is that they move in steps and thus introduce a noise in the system with a frequency that mainly depends on the velocity of the motion. In order to limit this issue, the RSU stepper motor is operated in micro-stepping mode and it is designed with a large inductance, reducing the maximum speed achievable by the motor but also smoothing the profile of the currents passing through its windings. For the RSU, the current in the motor winding is planned to be controlled in a closed loop, assuming as a reference a customizable current profile that can be updated even once in orbit.

We are also studying different options for the motor actuation profile and duration (between 8 and 15 s) in order to further minimize the generation of any additional micro-vibrations. To this purpose, a dedicated model has been developed to accurately predict the de-pointing and the micro-vibrations produced by the motions of the RSU by using MATLAB® and SIMULINK® (a block diagram environment for dynamic simulations). The model takes into account:

- The closed-loop current and the micro-stepping parameters of the motor control;
- The resistance, inductance, detent torque, back electromotive force constant, torque constant, and hysteresis torque factor of the motor;
- The gear ratio, backlash, variable mesh stiffness, friction and damping coefficients of the spur gear and pinion;
- The friction coefficients of the ball bearings;
- The inertia tensor (worst case) of the LS and FwS;
- The LS rigid-body and flexible modes and modal damping
- The LS and FwS rotation axes misalignments.

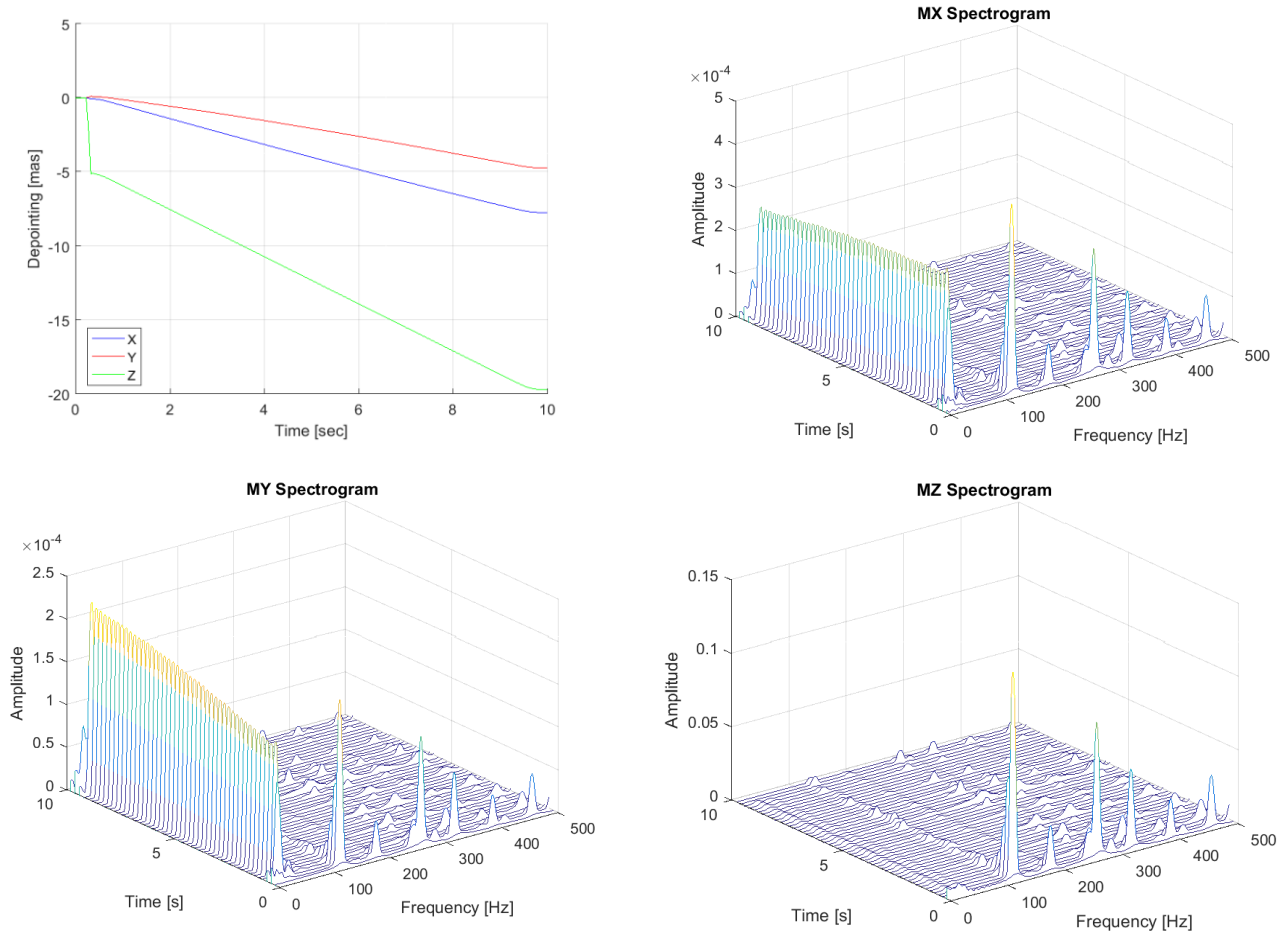

**Figure 3-1 Top left: predicted de-pointing induced by the RSU motion on the Euclid spacecraft. Top right: spectrogram of the exported torque on the x-axis. Bottom left: spectrogram of the exported torque on the y-axis. Bottom right: spectrogram of the exported torque on the z-axis.**

The LS flexible mode has been implemented in the micro-vibration model as follows. For each mode of the LS identified during modal analysis up to 500Hz (which is the highest frequency considered in the current version of the RSU micro-vibrations requirement), a frequency response function is generated, taking as input a torque and providing as output an angular displacement. The outputs of the micro-vibration model are expressed in angular de-pointing and exported torque which corresponds to the derivative of the angular momentum. The de-pointing and exported torque spectrograms predicted by the model are shown in Figure 3-1. The impact of the gear backlash onto the de-pointing on the z-axis is clearly visible at the beginning of the motion. It corresponds to about 25% of the total de-pointing.

As expected, the highest exported torque value is reached on the z-axis during the initial stages of the RSU motion, when most of the LS structural modes are excited due to the shock generated by the contact between the pinion and the gear (the frequency of the maximum value of the exported torque corresponds to the LS $1^{st}$ structural mode). The mismatch in the inertia of the LS and the FwS also contributes to enhance the peak value of the exported torque on this axis. On the x and y axes, the amplitude of the exported torque is significantly lower and virtually constant during the entire RSU motion. The main frequency at which the higher exported torque is measured corresponds for these two axes to the natural frequency of the motor. This mode is excited mainly due to the misalignment of the rotation axes of the LS and FwS. A similar peak is also present around the z-axis, but it is too low to be appreciated on the different scale of the corresponding plot. The current de-pointing and exported torque predictions induced by the RSU meet the unit requirements. The estimated RSU exported torque will be validated through direct measurements performed by the end of June 2016 on the RSU bread-board model (see Sect. 5), which is equipped with a flight representative LS and FwS.

## 4. MOST RECENT ISSUES

### 4.1 HDRM

Following an ESA alert concerning the type of magnet used inside the DC motor foreseen for the actuation of the HDRM, the inclusion of this sub-assembly in the RSU baseline design has been put into question. A new trade-off has been opened, as the LS static balancing was largely improved since the earlier design phases and the usefulness of the HDRM has now become questionable. Analyses demonstrated that no significant differences are expected in the LS displacements during launch with or without the HDRM and thus this functionality of the RSU can be most likely removed from the design. The outcomes of this trade-off will have to be consolidated and confirmed through the dedicated vibration tests of the RSU bread-board model (see Sect. 5), before the decision can be finally implemented into the flight design of the unit.

### 4.2 The RSU micro-switches

The micro-switch 11HM30-REL-PGM combined with the roller JS-151 manufactured by Honeywell has been selected as the baseline for usage within the RSU. These micro-switches are presently qualified to sustain <100'000 actuations in space conditions and only down to 218 K. An extended qualification of the micro-switches and the rollers is thus necessary before they can be safely implemented in the RSU design. This qualification is currently on-going, but preliminarily results revealed a failure of the switches after about $1/5^{th}$ of the total foreseen qualification cycles. The failure seems to be more related to the roller than to the switches themselves. It is thus likely that the micro-switches positioning close to the far end of the leaf shaft and the actuation strategy will be revised before the RSU design can be finalized for the flight model.

## 5. MODEL PHILOSOPHY

The entire RSU development program foresees the production of the following different models: the structural and thermal model (STM), the electrical model (EM), the bread-board model (BM), the qualification model (QM), the flight model (FM) and the flight spare (FS). In order to de-risk the development of the most critical parts connected to the RSU main drive train, APCO is also developing a drive train model (DTM) for internal test purposes that will not be described in detail here.

The RSU EM has been designed to mainly validate all electrical functions and interfaces of the units, verifying its ESD robustness and EMC properties. For this reason, all electronic components of the EM, except the mechanical parts, have been procured to comply with the established flights standard. The harness is representative of the FM design in terms of lengths, wire gauges, and shielding. A sketch of the EM and a picture of the real model are shown in Figure 5-1. The foreseen ESD and EMC tests of the EM were successfully passed at the end of 2015.

The RSU STM is a simplified model designed to be representative of the FM predicted mass and moment of inertia (MoI) and being used for the structural and thermal tests. The differences between the RSU STM characteristics and those expected for the final FM are that: (i) the primary stepper motor is replaced by two mass dummies; (ii) the HDRM, harness (including connectors) and the micro-switches are replaced by mass dummies; (iii) the ball bearings are replaced by standard industrial units with representative preload and stiffness; (iv) the flushing cover, encapsulating the main gear and used to avoid the dispersion of contaminants during operations, is replaced by a stainless steel mass dummy. All other elements are fully representative of the foreseen FM design, including materials and surface thermal properties. The STM has already passed the vibration and thermal tests. It will be delivered to undergo system level tests at the end of May 2016. A sketch of the STM and a picture of the real model are shown in Figure 5-1.

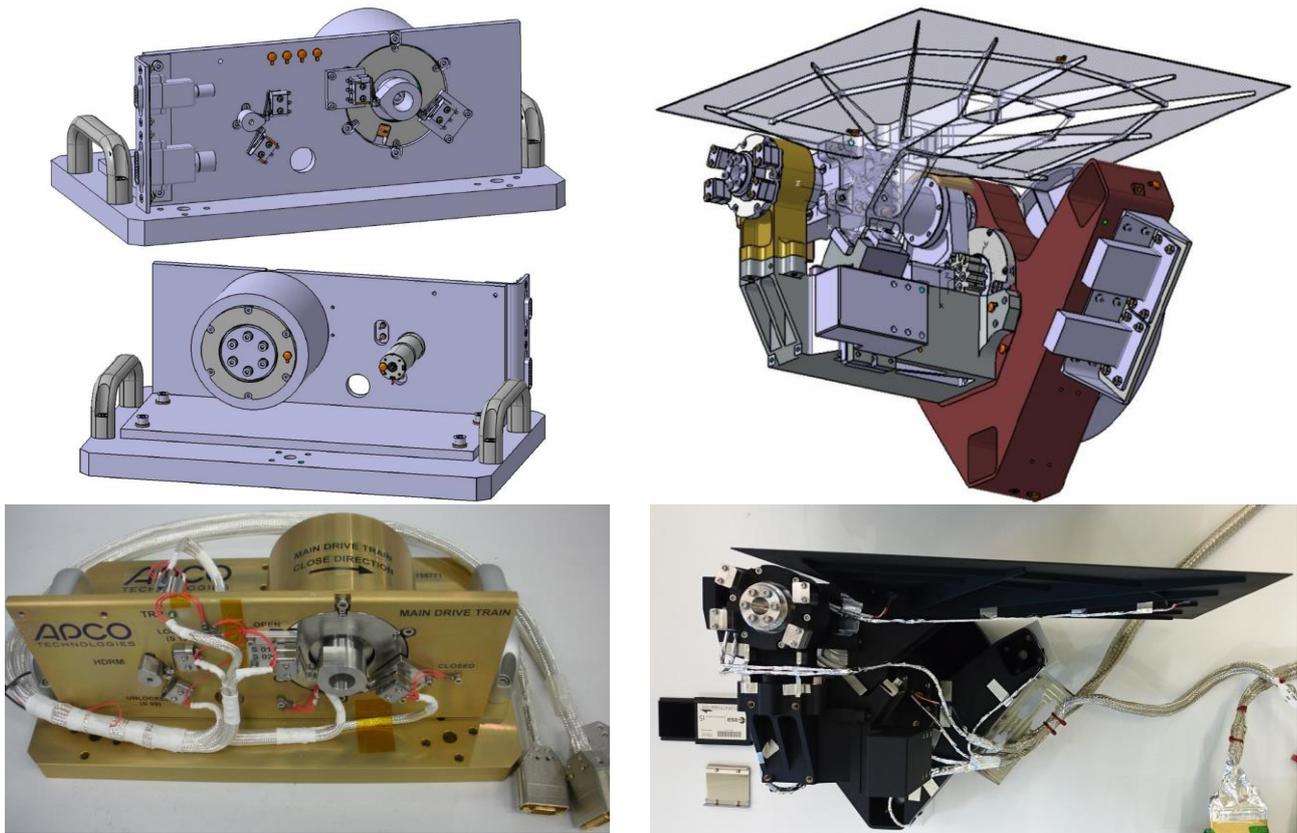

**Figure 5-1 Top and bottom left: a sketch and a picture of the RSU EM. Top and bottom right: a sketch and a picture of the RSU STM. The STM has been painted in black before undergoing the thermal test for thermal reasons.**

The main objectives of the RSU BM are the first measurements of the micro-vibration and exported torque performances of the RSU so far evaluated only through numerical modeling, and the first life-time test of the unit. In order to fulfill these objectives, the RSU BM has to include all micro-disturbance sources identified by previous analyses and requires LS and FwS to be fully representative of those of the FM in terms of materials, geometry, lubrication, assembly, adjustment and balancing procedure. Some geometrical simplification of the support structure is, however, applied for programmatic reasons. At the time of writing, the BM is being assembled at APCO and will undergo the micro-vibration and life-time test in the summer of 2016.

The Qualification Model of the RSU will be used for the full qualification of the mechanism. By definition, the QM is fully representative of the FM design and its manufacture is authorized by ESA only after a successful CDR. The QM will undergo several tests at qualification level, including a similar vibration and life-time tests as for the BM. At present, it is expected that the QM will be ready for tests in late 2016.

The FM will be fabricated and tested following the detailed design approved at CDR. It is presently expected that the manufacturing of the FM will be completed in mid-2017. A FS will also be produced and will be ready toward the end of 2017.

## 6. CONCLUSIONS

The present concept of the RSU has been developed and continuously improved in the past 5 years in order to comply with all challenging Euclid and VIS instrument requirements. The RSU EM and STM programs have now been completed and the two models successfully passed all foreseen tests. The present year is particularly crucial for the RSU team, as the unit will have to go through the CDR and perform the more demanding tests of the QM, freezing the FM design. The current schedule for all remaining RSU activities matches well the general Euclid/VIS milestones and we expect to be able to deliver the RSU FM by mid-2017.

## ACKNOWLEDGEMENT

The Euclid team at the University of Geneva and APCO acknowledge the support of the Swiss State Secretariat for Education, Research and Innovation SERI and ESA's PRODEX programme.